# A Contextual Model Of The Secessionist Rebellion in Eastern Ukraine

Olga Nicoara[†]
David White[*]

(Draft May 31, 2016)

## Abstract

This paper explores the possible contextual factors that drove some individuals to lead, and others to join the pro-secessionist rebellion in the 2013-2014 conflict in Eastern Ukraine. We expand on the existing rational choice literature on revolutionary participation and rebellious movements by building a contextual choice model accounting for both cost-benefit and behavioral considerations taken by Pro-Russian militants and rebels in the region of Donbass. Our model generates predictions about the characteristics of the socio-political-cultural context that are most likely to ignite and sustain hierarchical rebel movements similar to those in Ukraine.

1. Introduction

Over the past decade, Russia has re-emerged as a world military force. Since his return to power, Vladimir Putin has made it clear that he intends his sphere of influence to include all ethnic Russians across the world, especially those inhabiting the countries of the former Soviet Bloc. Ukraine has become the latest testing ground for the Russian geopolitical rhetoric. Pressures from both east and west have led to the armed conflict between pro-Kiev and pro-Russian forces, commonly known as the War in Donbass, which has dominated the news over the past two years. This conflict is shaping up to be a defining event in the relationship between Putin's Russia and the West.

[†] Email: olganicoara@gmail.com. Address: Department of Business and Economics, Ursinus College, 601 E. Main Street, 100 West College St., Collegeville, PA 19426.
[*] Email: whiteda@denison.edu Address: Department of Mathematics and Computer Science, Denison University, 100 West College St., Granville, OH 43023.

The ongoing war in Ukraine, has been preceded by similar separatist movements backed by Russian military forces: in Transnistria (Moldova, 1992), and in South Ossetia (Georgia, 2008), resulting in the existing, so called "frozen conflicts". If our goal is to predict and attenuate future cases of global violence, the key task is to identify and distinguish all the possible causes for conflict. Given Moscow's involvement in Ukraine—as reported by the media and by the Organization for Security and Cooperation (OSCE)—two questions arise. First, how much of the pro-separation conflict in Eastern Ukraine arose as a result of actual democracy demands by the Russian ethnics? Second, how much of the conflict was induced or provoked by Russian political propaganda and deliberate interference with the incentives faced by the population? Compared to past cases of breakaway wars in other former Soviet republics (e.g., Moldova and Georgia), the War in Donbass stands out in its military force and violence. What factors influenced the decision of individuals of ethnically and politically distinct groups to start and fight this violent secession war in Ukraine?

What was the logic of collective action in the Ukrainian war? The rational choice literature accounts for the manifestation of various rebellious events and outcomes in history (see Mahoney 2000). It assumes that rebellions are a manifestation of uneasiness with a dominant regime, and that most often individuals in repressed societies fail to start a revolution due to inherent collective action problems that plague large groups of individuals even if they would jointly benefit from it (Olson

1965, Ireland 1967, Tullock 1971, Tullock 1974, Lichbach 1994, 1995, Leeson 2010). What stops rebellions from occurring from a pure cost-benefit perspective are the disincentives created by the high costs of individual participation in and leading of a revolution. As rebellions require large groups for success, they produce larger incentives for individuals to free ride on the participation and leadership acts of others. Under the free rider's premise, historically, one way small, rebellious groups were able to overcome the incentive problems was by devising a diversity of institutional solutions that improved the cost-benefit balance of individual participation, leadership, and ultimately successful collective action (Lichbach 1994, 1995, Leeson 2010). Timur Kuran theorized that individuals in communist societies deliberately falsified their (true) preferences about the regime, up to a critical point that eventually triggered, from "sparks" to "prairie fires", the mass protests and revolutions that took place in Eastern Europe towards the end of communism (Kuran 1989).

Several other studies include the role of socio-political-cultural context in the rise of secessionist movements in the world and in the former Soviet space (see, for example, Pavkovic and Radan 2013; Coppieters and Sawka 2003). In this paper, we extend the existing contextual studies of and the existing solutions to revolutionary non/participation (see, for example, Tilly 1978; Kuran 1989, 1991; Moore 1995; Kuran 1989, 1991; Lichbach 1994, 1995; Kurrild-Klitgaard 1997; Petersen 2001; Leeson 2010) and violent secessionist movements (Moore 1998, Hale 2000), by building a contextual choice model that accounts for both behavioral and pure cost-benefit considerations[1] taken by the individuals in the different groups of militants and rebels in Donbass, Eastern Ukraine.

Our model extends the theoretical approaches in Tullock (1971, later 1974), Kuran (1989, 1991), Lichbach (1994, 1995), and Leeson (2010), and adds a term to factor in cultural beliefs and values. The contextual feature of our model of choice generates predictions about the characteristics of the socio-political-cultural contexts that are most likely to ignite and sustain violent rebel groups similar to those in Ukraine.

We focus solely on the contextual aspects of choice of individuals in Donbass, and our intention is to understand the effects of history, politics, and culture in their calculations of participation or leadership of protests or rebellions. We then discuss how these key factors might have resulted in frozen conflict areas in former Soviet Bloc countries in the past, and what steps might be taken to avoid replicating the conflict in Ukraine in the future. Due to our focus on Donbass and Russia, we have not included a discussion of the motivations of Euromaidan protesters in Kiev or of Russian supporting residents of Crimea. We do, however, demonstrate how these events outside of Donbass affected the budding rebellion.

2. **Background on the Secessionist Movement in Eastern Ukraine**

In this section, we present the chronology of events in Donbass from the early part of 2014. These events demonstrate that there were steep costs for individuals

---

[1] In Roger Petersen's approach to explaining individual rebellious action: "Whether individuals come to act as rebels or collaborators, killers or victims, heroes or cowards during times of upheaval is largely determined by the nature of their everyday economic, social, and political life, both in the time of the upheaval and the period prior to it. The extraordinary is inextricably linked to the ordinary." (Petersen 2001)

(e.g. pro-Kiev demonstrators) acting against the goals of the Pro-Russian militants, and benefits to individuals acting in favor of the pro-Russian movement. These events also demonstrate the role Russian propaganda played in convincing members of the pro-Russian movement to act on their beliefs (e.g. via convincing them that they would be supported by Russia).

Russian propaganda has been present in Donbass for the past several years, and its volume increased when Vladimir Putin took over media outlets and began his program of calling on ethnic Russians everywhere (Sonne, 2014). Freedom House, a watchdog organization devoted to promoting freedom, maintains information about media outlets across the world. They report (2014) that the Russian government restricts international radio and television broadcasting. In particular, the restrictions are on private FM radio stations that rebroadcast news from the British Broadcasting Corporation, Radio Free Europe/Radio Liberty (RFE/RL)[2]. This demonstrates the extent to which Putin controls the Russian media. Freedom House (2015) also reports that Russian propaganda is broadcast into Ukraine, and that these broadcasts frequently contain false information designed to convince Ukrainian citizens that they are being oppressed by the Ukrainian government.

A thorough analysis of Russian sentiment was conducted by Andrew Wilson (Wilson, 2014). He found that Russia has long viewed NATO as encroaching on Russian territory, and that Russians feel surrounded, with an increasingly united

---

[2] Freedom House informs that: "in November 2012, RFE/RL lost its medium-wave local broadcasting license due to the implementation of a 2011 law prohibiting foreign ownership of broadcast media." (https://freedomhouse.org/report/freedom-press/2013/russia#.VSa1ykgl5Qc )

Europe on one side and the United States on the other. Russia views the European Union (EU) as an extension of NATO and therefore wishes to prevent Ukraine from joining the EU, so that Ukraine can serve as a buffer between Russia and NATO. For this reason, Russia pressured Ukraine's president Viktor Yanukovych to accept a bailout in November 2013 in return for his promise to not join the EU. Shortly thereafter the Euromaidan movement in Maidan Square in Kiev began to protest. They protested throughout the early part of 2014. On February 23, Yanukovych was ousted from power and fled to Russia.

Beginning February 26, Russia sent troops into Crimea. By early March, these troops controlled Crimea. An ad hoc referendum, on whether Crimea should formally join Russia, was held on March 16, and a majority voted in favor. The reliability of this vote has been questioned (European Commission, 2014). Meanwhile, unrest grew throughout February and March in Donbass, as ethnic Russians raised on the propaganda that they were being oppressed, mistreated, and embarrassed by Kiev (Freedom House, 2015) began to advocate for their rights. For example, ethnic Russians in Donbass began demonstrating in favor of making Russian a formal second language of Ukraine, and demonstrating against the Euromaidan demonstrators in Kiev (i.e. against joining the EU). As unfolding events made it clear that Kiev's government was weak (e.g. unable to prevent the referendum in Crimea), these demonstrations occurred with increasing frequency and strength. Demonstrations were strongest in the cities of Donetsk and Luhansk, two of the most populous cities in Donbass. Demonstrators began rioting and making demands on March 1. The first major, violent act of the budding rebellion was when rioters took over the Donetsk Regional State Administration (RSA) Building during the period of March 1-6.

While these rioters were making their decisions, they were hearing (primarily from Russian news outlets) about Russia's move to take Crimea, about Ukraine's weakness in the capital, and about how other media should not be trusted (e.g. was being manipulated by NATO). These factors fed into the decision of the rioters to become rebels seeking independence from Kiev. On March 25, the Ukrainian government banned four Russian television channels in Donbass, claiming that the broadcasts "threaten the national security, sovereignty, and territorial integrity of Ukraine" (Kates, 2014), but this action came too late, as the rebellion was already underway. Even the pro-Russian rebels realized the value of Russian broadcasts: in April, when the offices of the state television network in Donetsk were captured, the rebels disabled Ukrainian broadcasts and began again broadcasting "Kremlin-backed Russian channels" (Tsvetkova, 2014).

Chronology Of The Main Events Leading to Conflict In Donbass:

- **March 1, 2014** - a group of pro-Kiev protesters are attacked and beaten by pro-Russian demonstrators in Kharkiv, Donbass. The police do not intervene. In our model, this increases the fear of punishment[3] by pro-Russian rebels.

- **March 11, 2014** - the Ukrainian National Council for TV and Radio Broadcasting ordered all cable operators to cease broadcasting a number of state controlled Russian channels. The National Security and Defense Council stated that these broadcasts were a threat to national security. Evidence was presented demonstrating that the Russian broadcasts contained staged, fake news designed to make Ukraine look bad (Ennis, 2014).

- **March 13, 2014** - a group of pro-Maidan protesters are beaten, and one is killed, by pro-Russian demonstrators in Donetsk, Donbass (Macdonald, 2014). This causes an increase in the fear of punishment, as above.

---
[3] Factor $C^i$ in our model.

- **March 16, 2014** – pro-Russian demonstrators hold staged rallies in favor of a referendum, in several cities in Donbass (Laughland, 2014). In our model this contributes to an increase in reputational gain for publicly supporting the pro-Russian movement, and hence contributes to the threshold effect which causes the rebellion to reach critical mass.

- **March 18, 2014** - Putin states that Russia will protect Russian-speaking people everywhere, including in Ukraine (Kendall, 2014). In our model, this lowers the perceived probability of arrest by Ukrainian forces and increases the probability of the rebellion succeeding.

- **March 25, 2014** - Ukraine bans Russian TV (Kates, 2014), proving that Russian propaganda was a significant enough influence for Kiev to actively try to block it.

- **April 6, 2014** – this date marks the true beginning of the War in Donbass. The Pro-Russian rebels announced a referendum and seized several government buildings. In response, the central Ukrainian government sent counter-terrorism forces, and increased fighting ensued.

- **Spring, 2014** – attacks against pro-Ukrainian protestors by pro-Russian rebels become increasingly common, contributing to increased fear of punishment in our model. Attacks occur in Luhansk (Fedosenko, 2014), Donetsk (Maceda and Kovalova, 2014), Slavyansk (Luhn, 2014). Furthermore, the city of Sloviansk, controlled by pro-Russian rebels, is besieged by Ukrainian armed forces, Donetsk is shelled, and Putin reiterates his promise to protect Russian speakers everywhere (Coalson, 2014). These events contribute to fear of punishment for pro-Ukrainian protesters, fear of reprisals by residents of Donbass (especially Sloviansk and Donetsk), and an increase in the perceived ability of pro-Russian forces to win[4].

Three main factors emerge from the chronology above and contribute in our model to shaping the context of the armed conflict and rebellion in Donbass:

1. <u>Propaganda through state ownership of media in Russia:</u>

    We argue that pro-Russian propaganda on local media in Donbass created beliefs that life under Russian, or Russian-like rule is better than life under the existing Ukrainian rule. The idea of a "New Russia" – otherwise an archaic

---

[4] Respectively, these three factors are $C^i$, $c^i$, and $p_R^i$ in our model.

term in Russian history encyclopedias – promoted by pro-Russian demonstrators and elsewhere on Russian TV and local radio, paints the picture of a wonderful life in Russia, and a superiority of Russians and the Russian form of governance and culture globally. Furthermore, Russian propaganda in 2013-2014 gave the impression that ethnic Russians in Ukraine (and elsewhere) belong with Russia (Freedom House, 2014, 2015). Lastly, the propaganda gave the impression that ethnic Russians in Ukraine were being oppressed by the Ukrainian central government, even if this impression is not borne out by the evidence.

2. <u>Fear of acting or speaking against the pro-Russian movement:</u>
    We argue that the violent reprisals of the pro-Russian forces against pro-Kiev demonstrators contributed substantially to fear factors in our model, leading to a decreasing willingness of Donbass residents to resist the slide of the region into violent rebellion. In our model these fear factors are present both for supporters of Kiev and for residents without strong preferences regarding the governance of Donbass.

3. <u>Consolidation of Putin's power following the Russians taking over Crimea:</u>
    By taking Crimea, Vladimir Putin gave residents of Donbass hope that they could also join Russia. The referendum in Crimea spawned a similar referendum in Donbass, despite pressure from Russia against the hosting of such a referendum.

**3. A Contextual Model of Choice**

Building on the existing literature modeling revolutionary non/participation (in particular, Kuran 1989, 1991; Moore 1995; Lichbach 1994, 1995; Petersen 2001; Leeson 2010), we create a contextual choice framework to account for the variability in individuals' perception of costs and benefits that led to the start of the War in Donbass.

We consider several different option groups: the pro-Russian rebels, the pro-Western movement, and the pro-Ukranian. We seek to understand how contextual factors, derived from the chronology of events outlined in the previous section, influenced the cost-benefit calculations that went into the decision of a resident in the Donbass region to either join or not join the armed rebellion.

For any individual *i* in Donbass, contemplating rebellion in the first half of 2014, the following factors come into play:

1. $F^i$ the value of freedom from Kiev. Note that different individuals may have different views on what this freedom will mean. For some it means joining Russia. For others it means creating an independent, legitimate, democratic state. For others it means seizing power to hold at gun point.
2. $S^i$ the value of life under the status quo, i.e. if Kiev regains power over Donbass.
3. $A_U^i$ the cost of being arrested if Kiev regains power and if the individual participated in the rebellion.
4. $A_R^i$ the cost of being arrested if the rebellion succeeds and the individual *i* actively opposed the rebellion.
5. $c^i$ the cost, inflicted by pro-Russian forces, of not joining the rebellion. This cost might include loss of social status, property loss, or even receipt of fewer rations in a city under siege.

6. $C^i$ the cost, inflicted by pro-Russian forces, of advocating in favor of Ukrainian rule in Donbass. This cost might include beatings, property loss, being forced into hiding, and even death.
7. $V^i$ the desire for/against violence.
8. $R(y^i)$ the increase/decrease in reputation for publicly taking position $y^i$. Here $y^i$ could be publicly supporting the pro-Russian rebellion or publicly supporting the pro-Ukrainian forces.
9. $N(y^i|x^i)$ the integrity for publicly taking position $y^i$ while privately supporting $x^i$. Here $x^i$ could be either pro-Russian or pro-Ukrainian. This term can be viewed as measuring the cost to individual *i* for publicly espousing a position in conflict with his/her private beliefs.

We shall assume all but the last three of the quantities listed above are positive, so our equations for expected payoffs will use subtraction for quantities where the payoff is negative (e.g. $A_U, A_R, c, C$). Our evidence for including $C^i$ and $c^i$ are the events chronicled in Section 2. We also tacitly assume $C^i > c^i$, which also explains our notation. The last two quantities listed above also appear in (Kuran, 1989). We allow the last three quantities $(V^i, R(y^i), N(y^i|x^i))$, to be either positive or negative[5] depending on the individual *i*. Neither we, nor Kuran, claim to have precise formulas for these three quantities. A first approximation to $R(y^i)$ might be $\alpha * \frac{\#S}{\#T}$ where $\alpha$ is some scaling factor, #S is the number of individual *i*'s friends supporting position $y^i$, and #T is the total number of friends of *i*. A weighted version of this formula, with friends weighted by how much individual *i* respects their opinions, would be better[6]. An even better approach would mimic Google's PageRank algorithm and conduct an iterative approach to fully include the network structure when computing the weights.

---

[5] Alternately we could also introduce a sign variable $\sigma_i$. Kuran assumes $0<y^i$ and $x^i<1$, but this convention does not affect the theory in any way.

Even without a precise formula for $R(y^i)$, some heuristics are clear. For example, as the number of public supporters of the pro-Russian movement increases, so does $R(y^i)$, where $y^i$ is the position of supporting the pro-Russian movement (similarly, the reputational gain of supporting Ukraine goes down).

### 3.1. Solving The Individual Participation Problem

Let $p_R^i$ be the probability of the rebellion succeeding, according to individual $i$. The expected payoff of joining the pro-Russian forces is then

$$E(R) = p_R^i * F^i - (1 - p_R^i)A_U^i + R(y^i) + N(y^i|x^i) + V^i$$

Here $V^i$ would be positive if individual $i$ enjoys violence and would be negative otherwise. We assume that joining the pro-Russian forces (or indeed, the pro-Kiev forces) entails a non-negative amount of violence, which can be zero. Likewise, $R(y^i)$ would be positive if the majority of the value that individual $i$ derives from reputation comes from pro-Russian individuals. Lastly, $N(y^i|x^i)$ would be positive if $y^i = x^i$ and negative otherwise since, as Kuran points out, there is a cost to the individual for preference falsification. The expected payoff for joining the pro-Ukrainian forces is:

$$E(U) = S^i(1 - p_R^i) - A_R^i * p_R^i - C^i + R(y^i) + N(y^i|x^i) + V^i$$

---

[6] Kuran also defines public sentiment as $\sum w_i y_i$, where $w_i$ is the weight of individual $i$'s decision. He then uses $y^i$ to argue about thresholds and domino effects.

Lastly, the expected payoff for not joining either side is:

$$E(NJ) = S^i(1 - p_R^i) - c^i + R(y^i) + N(y^i|x^i) + V^i$$

Here $V^i$ will be negative if the individual wants violence and positive if the individual does not like violence. We expect $R(y^i)$ to be fairly small, though perhaps larger than the reputation gained from joining the pro-Russian rebellion, depending on the social network of individual $i$. Note that $N(y^i|x^i)$ could be negative, either if the individual's preference $x^i$ is to advocate for remaining with Ukraine but the individual does not (perhaps $i$ is afraid of the reprisal $C^i$), or if the preference is pro-Russian but the individual does not join the pro-Russian forces (perhaps the individual is afraid of arrest $A_U^i$).

Let us consider some heuristics for the quantities in the equations above. We encourage the reader to refer back to the chronology of events for our justification of these heuristics. First, as the number of Russian troops in Donbass increased, the cost $C^i$ increased, and this resulted in fewer and fewer pro-Ukraine demonstrations. Similarly, the probability of arrest by the police decreased, both because the police cannot arrest a critical mass, but also because the police might be afraid to confront armed, trained Russian soldiers. Additionally, after Crimea's take-over by Russian forces, $F^i$ increased because many in Donbass began to believe they might also join Russia (as demonstrated by the demonstrations in favor of a referendum to join Russia, and the referendum itself in April). Even for an individual $i$ who wanted freedom from Kiev (but perhaps did not want to join Russia) there would be an increase in $F^i$ after $i$ realized Kiev's weakness and his/her ability to profit in the new power vacuum.

As the amount of Russian propaganda streaming across the border from Russia to Donbass increased, the perceived probability of success $p_R^i$ for the rebellion increased (because more and more individuals perceived that Putin would help them if they needed his help). Similarly, the perceived benefit of $F^i$ increased because the propaganda and referendum made it appear that joining Russia was a real and positive possibility. Similarly, the perceived payoff of remaining under the status quo $S^i$ decreased because the propaganda helped convince some that they would be better off without Ukraine, or even that they were being oppressed by Ukraine (Freedom House, 2015). Lastly, the reputational benefit from joining the rebellion increased, as we have already remarked.

### 3.2. Solving The First-Mover Problem

We now consider the first-mover problem for individuals where true preference $x^i$ is pro-Russian. For such individuals, $p_R^i$ must have seemed fairly small. However, personal beliefs about how much better life would be without rule from Kiev (perhaps influenced by Russian propaganda) can overcome this small $p_R^i$. More importantly, $R(y^i)$ can be strongly influenced by cultural beliefs, as evidenced through historical research on "Russian conservatism" by Richard Pipes (Pipes, 2005), and manifested by an increase in pro-Russian demonstrations in Moscow and in Kiev over recent years. Ariel Cohen (Cohen, 2007) observes that, since Putin's consolidation of power in the Kremlin, "ethnic nationalism and extremism have reemerged in modern Russia", while the government has adopted authoritarian tendencies and has attempted to imbue in citizens a deep trust in the government and a belief in the superiority of the Russian form of governance. As a result, citizens and have been seen to stand up, defensively, to non-Russian authority increasingly in recent years. In a related vein, Andrew Wilson discusses "Russia's addiction to dangerous myths," and claims that Russians have "the world's biggest persecution complex" (Wilson, 2014). This suggests that, for pro-Russian individuals, the effect of $A_U^i$ may not be as large as an observer might believe, e.g. because being arrested while standing up to authority could give even more positive feedback to one's reputation $R(y^i)$.

Lastly, individuals with lexicographic preferences for violence will have $V^i$ large and positive, and this will help offset the large negative term $(1 - p_R^i)A_U^i$. We therefore conclude that first-movers will be drawn from intrinsically violent individuals, individuals who feel $p_R^i$ is not so small (e.g. who believe Russian forces will swoop in and aid the rebellion), individuals with strong anti-authority (and pro-Russian) culture, and individuals who feel the benefits of a successful rebellion ($F^i$) are so much greater than the current status quo ($S^i$) to justify the risk of arrest $A_U^i$.

After these first-movers start the rebellion, the proportion of publicly pro-Russian supporters increases. This causes a threshold phenomenon, similar to that observed by Kuran, wherein the next wave of slightly more cautious individuals join the rebellion, leading to an increase in the number of pro-Russian supporters, leading the next group of even more cautious individuals to join, and so on, until a majority of individuals have joined the rebellion. Our model shows that, after enough pro-Russian supporters have joined the movement, others who want independence (but not necessarily to join Russia will join), because they see that Ukraine is weak and now is the time to break away. Lastly, in the face of a majority in favor of the movement, individuals who fear the costs $C^i$, $c^i$, and $A_R^i$ of not joining or of resisting the rebellion will publicly support it out of fear.

For individuals whose true preference is pro-Ukrainian, the default position is to allow the police to stop the rebellion. After realizing that the police are incapable, individuals whose integrity demands action might decide to protest against the pro-Russians. As discussed in Section 2, these protests occurred frequently in the early days of the rebellion. As the chronology of events demonstrates, both the pro-Russian and pro-Ukrainian forces were violent, so again individuals with a lexicographic preference for violence would be likely first-movers. Clearly, if the expected payoff for joining the pro-Ukraine movement is to be positive, then $S^i(1-p_R^i), C^i$, and $N^i$ would need to be greater than $C^i, p_R^i$, and $A_R^i$. However, as discussed in Section 2, when the rebellion got stronger, the cost $C^i$ of opposing it increased. Time and again, pro-Russian mobs attacked pro-Ukrainian demonstrations with baseball bats (see chronology above), demonstrating to all that $C^i$ could be large (and suggesting that the cost $A_R^i$, after a rebel victory, might be even more brutal). Meanwhile, $p_R$ increased (from all individuals' point of view) as the rebellion grew. Our model predicts in this case that the number of individuals publicly supporting Ukraine would decrease under these circumstances, and indeed that is precisely what occurred in the late spring and early summer months.

In the next section we discuss the dynamics of the three populations: pro-Russian, pro-Ukrainian, and those who do not advocate for either position. Our model allows individuals to flow back and forth between these groups, and we study how the various factors in our model drive these types of decisions. Observe that the system is not closed: some individuals in Donbass (e.g. pro-Ukrainian activists) may choose to flee back into conflict-free Ukrainian territory rather than continue living under pro-Russian rule. Some individuals who are pro-Russian but do not wish to join the rebellion might choose to simply move to Russia. Lastly, individuals in the third group might choose to flee the region to avoid the conflict altogether, regardless of their preferences for who will rule in Donbass. Indeed, a recent report by The Office of the United Nations High Commissioner for Human Rights (OHCHR) suggests that upwards of one million people chose to flee the region, either internally (within Ukraine), or externally (mostly in Russia), rather than join any of the groups of actors we have considered. Note in addition that individuals from Ukraine (e.g. members of Ukraine's state security service, the SBU) can enter the system and decrease the probability of the rebellion succeeding, and individuals from Russia (e.g. the `Little Green Men' stationed in Crimea in Spring 2014[7]) can enter the region and increase the probability that the rebellion succeeds. We believe these groups merit their own study, and that they would be a good subject for future work, but we have not focused on them here.

**3.3. Comparing The Payoff Equations**

We now analyze the payoff equations to determine the dynamics of the system. We first consider the decision of not joining either group (NJ) vs. the decision to advocate in favor of Ukraine. Our analysis of events in 2014 suggests that, as time went on, the payoff for the former was greater than the payoff for the latter, leading to fewer and fewer pro-Ukrainian protests. We write this inequality in the language of our model, introducing subscripts to help us remember if the *R*, *N*, and *V* terms are for the NJ population or the U population:

$$S^i(1-p_R^i) - c^i + R(NJ) + N(NJ|x^i) + V_{NJ}^i >$$
$$> S^i(1-p_R^i) - A_R^i * p_R^i - C^i + R(U) + N(U|x^i) + V_U^i$$

---

[7] Richard Sawka writes of "some 12,500 service people stationed in the peninsula … wearing green uniforms without insignia, the men claimed to be local volunteers and were soon dubbed 'little green men'. In fact, they were highly trained Russian special forces using advanced technologies to achieve the bloodless takeover of the peninsula." (Sakwa 2015, p.104)

After removing like terms and reshuffling terms we are left with

$$[R(NJ) - R(U)] + [N(NJ|x^i) - N(U|x^i)] + [V_{NJ}^i - V_U^i] > -A_R^i * p_R^i + [c^i - C^i]$$

As we have remarked, $c^i - C^i$ is negative, and beatings by pro-Russian mobs only make it more negative, as well as making -$A_R^i$ more negative. The inequality is therefore satisfied as soon as the left-hand side is positive, or at least not too negative. As more and more individuals support the rebellion, the first term becomes positive rather than negative, since supporting Ukraine will be more and more likely to lead to negative reputational benefit. Next consider the *V* terms: we assume preference towards violence is an intrinsic quality and therefore time independent. It is possible individual *i* has a large positive $V_U^i$ (i.e. prefers violent settings), causing the third term on the left above to be negative. Even so, this difference is constant, whereas the right hand side is increasingly negative as time passes. Therefore, even pro-Ukrainians who prefer violence will eventually decide it's better not to oppose the pro-Russian forces. Lastly, we consider the integrity terms *N*. These terms seem most likely to force the left-hand side above to become negative, since if $x^i = U$ then $N(NJ|x^i)$ is negative and $N(U|x^i)$ is large and positive, so the middle term on the left above is large and negative. The relationship between integrity and time is unclear; Kuran points out that $N(y|x^i)$ is increasingly negative over time if $y^i \neq x^i$, but it is not clear that $N(x^i|x^i)$ is increasingly positive over time. Furthermore, the weight of the integrity term relative to the other terms above has not been carefully studied, partially due to the difficulty of obtaining data on this question. However, the increase of $C^i$, $A_R^i$, and $p_R^i$ over time is effectively unbounded (especially considering that several pro-Ukrainian protesters died), whereas integrity terms are most likely bounded, since acting against one's preferences is usually preferable to beatings or death. This explains why fewer and fewer would choose to publicly support the pro-Ukrainian movement over time.

For a different derivation of this fact, consider solving the equation above for $p_R^i$. Then an individual will choose $y^i = NJ$ over $y^i = U$ as soon as:

$$p_R^i > \frac{c^i - C^i - R(NJ) + R(U) - N(NJ|x^i) + N(U|x^i) - V_{NJ}^i + V_U^i}{A_R^i}$$

Viewed in this light, we see that the strictly increasing nature of $p_R^i$ (making the left-hand side larger), of $A_R^i$ (making the right-hand side smaller), and of $C^i$ (making the right-hand side smaller) contribute significantly to this equation being satisfied over time, hence to individuals choosing to do nothing (NJ) rather than demonstrate in favor of Ukraine (U). Furthermore, as the most fearful pro-Ukrainians allow themselves to be silenced, $p_R^i$ increases even more, leading to a Kuran-style threshold in which increasing numbers of pro-Ukrainians are silenced until the only supporters remaining are those for whom the integrity term dominates all others[8]. Thus, we have explained the decrease in pro-Ukrainian protests—matching observations from 2014—due in large part to attacks on pro-Ukrainian protesters gathered at rallies in Donetsk (*Reuters* 2014).

We turn now to the question of whether an individual *i* chooses to publicly support the pro-Russian movement ($y^i = R$) or whether to do nothing ($y^i = NJ$). We choose to omit the case of an individual deciding between pro-Russian and pro-Ukrainian activism, because these movements were in direct opposition to one another, making the third option (NJ) more attractive. In order for an individual to choose R over NJ the following inequality must be satisfied:

$$p_R^i * F^i - (1 - p_R^i) + R(R) + N(R|x^i) + V_R^i$$
$$> S^i(1 - p_R^i) - c^i + R(NJ) + N(NJ|x^i) + V_{NJ}^i$$

Solving for $p_R^i$ we obtain the following equivalent inequality:

$$p_R^i > \frac{S^i - c^i + A_U^i + r(nj) - r(r) + N(NJ|x^i) - N(R|x^i) + V_{NJ}^I - V_R^i}{F^i + S^i + A_U^i}$$

We use this inequality to make several observations. First, if the perceived value $F^i$ of independence from Kiev increases (e.g. due to Russian propaganda), then the right hand side decreases, making *i* more likely to support the rebellion. As the cost $c^i$ of not joining increases (e.g. due to pro-Russian forces gaining more power in Donbass), then the right side decreases, making *i* more likely to support the rebellion. Next, consider the reputation terms. Observe that R(NJ) can be assumed small because one does not derive reputational gains from taking no action. While R(R) may have been negative at the start of the rebellion; it became increasingly positive as the number of supporters of the rebellion increased. This shift causes the right hand side of the inequality to decrease, making $i$ more likely to support the rebellion.

---
[8] One could similarly solve the equation above for $C^i$ or $A_R^i$, and would discover the same conclusions due to the same considerations.

Next, consider the integrity terms. If an individual internally supports the pro-Russian movement then this decreases the right-hand side of the inequality. If an individual does not support the pro-Russian movement, then $N(R(x^i))$ is negative, so this increases the right-hand side and makes it less likely $i$ would support the rebellion. However, as we argued in the previous comparison, the integrity terms are bounded and can be dominated by fear terms such as $c^i$.

If the preference for violence is large then the right hand side decreases, making $i$ more likely to support the rebellion. Conversely, if the desire against violence is large, so $V_{NJ}^i > 0$ and $V_R^i < 0$, then the individual would be less likely to support the pro-Russian movement. However, as discussed above, the violence preference term is constant over time, hence can be dominated by the other terms in the inequality over time.

Lastly, we consider the terms appearing in both the numerator and the denominator. In general, fear of arrest would be a large motivating factor against rebelling. However, as time passed and evidence accrued to demonstrate that most pro-Russian activists were not being arrested, $A_U^i$ decreased. In the fraction above, this has the effect of allowing the other terms to become more important in the decision to rebel or not rebel. Similarly, as propaganda made the status quo $S^i$ appear less attractive, that term decreased, allowing terms such as $C^i, c^i$ and $R(R)$ to play a larger role. Finally, while all of the terms above were shifting, so too was $p_R^i$ increasing, e.g. as more people publicly supported the rebellion, as fewer protesters were being arrested, etc. This makes the left-hand side larger, so the inequality is more likely to be satisfied.

To summarize, pro-Russian propaganda, an increasing number of public supporters of the pro-Russian movement, an increasing fear of the consequences $c^i$ of not joining the pro-Russian movement, and a decreasing fear of arrest all contributed to making the inequality above easier to satisfy, and therefore making individuals more likely to join the rebellion.

Model shortcomings:

We only consider actors in Donbass. Our model includes pro-Russian rebels, pro-Ukraine activists, and residents of Donbass who do not take action in favor of either group. Our model includes factors for preference falsification following Kuran, and these terms are particularly important for the third group of actors listed above. Our model also factors in individuals with a lexicographic preference for violence, and we demonstrate that these individuals will be the first-movers.

Russians in Russia are not explicitly present in our model. However, Russian propaganda and the existence of Russian troops in Donbass (whether they were there officially or not) strengthen the belief in success of the first group of actors we consider and help push them towards action. The induced beliefs[9] of the pro-

---

[9] We call "induced beliefs" the beliefs individuals come to hold as a result of their persisted exposure to censored media channels (used as a tool to influence popular opinion in against political threats and/or rivals).

Russian rebels explain their willingness to be first-movers, as well as why the rebellion spread so quickly in the spring of 2014.

A more complex model could also feature Ukrainian peace keeping forces in Donbass, Russian forces in Donbass, and Russians in Russia who support Donbass's pro-Russian movement. Similarly, it would be worthwhile to consider how the Euromaidan movement fed into the start of the rebellion in Donbass. Euromaidan supporters were in general anti-Russian, and indeed were protesting partially against President Yanukovych's deal with Russia to keep Ukraine out of the European Union (BBC, 2014). Ironically, their protests in Kiev forced Yanukovych out of power, and consequently demonstrated the weakness of central Ukrainian authority, perhaps contributing to the budding rebellion in Donbass. However, since these actions did not take place in Donbass, we have chosen not to include them in our model. We therefore leave such considerations to the interested reader.

**4. Conclusion**

In this paper we built a model which factors in rational decision-making, lexicographic preferences for violence, preference falsification for non-activist residents, and the cultural beliefs of pro-Russian rebels.

We have argued that this specific cocktail of factors led to the start of the War in Donbass and explains the speed with which the rebellion spread. If, in the future, similar factors are identified in other post-Soviet bloc countries, then the same model (perhaps with different cultural factors) would suggest rebellion to soon follow. Given Vladimir Putin's rhetoric about ethnic Russians worldwide (see Coalson, 2014), and given the continuing Kremlin control of Russian TV and radio, including broadcasts into neighboring countries (Freedom House, 2015), it is very possible these factors could be repeated in the near future. If the EU and NATO wish to prevent the War in Donbass from being repeated, then steps must be taken to reduce the effect of the factors which led to rebellion in Ukraine.

For example, pro-Russian propaganda should be limited. The Ukrainian government took this step in Donbass (Kates, 2014, Prentice, 2014), but it was too late to have any effect. Furthermore, steps must be taken to convince ethnic Russians that they are not being persecuted by the local or national government. As stated in the introduction to Andrew Wilson's book, ethnic Russians have "the world's biggest persecution complex." It is important, therefore, to factor this cultural observation into all interactions with ethnic Russians.

Another important lesson from the War in Donbass is the importance of protecting individuals who are not part of the pro-Russian movement. In most democratic Western countries protesters from both sides can safely demonstrate without fear of reprisal. The evidence we have presented demonstrates conclusively that such was not the case in Donbass in the spring of 2014. Pro-Ukrainian demonstrators were attacked time and again, and the police did little to protect them. Our model demonstrates how this drove preference falsification for individuals from the pro-Ukraine movement, and the resulting asymmetry produced the appearance of more support for the pro-Russian movement than was actually present. A domino effect, similar to that predicted by Kuran, followed. In order to prevent this sequence of events from recurring in other post-Soviet bloc countries, efforts should be taken to retain police officers who support the national government, and to restrict the number of pro-Russian police officers as much as possible without feeding into the persecution complex of local ethnic Russians.

Lastly, the cultural effects in our model suggest that the West would do best to limit their role in Ukraine and similar countries. This conclusion is in opposition to the call for intervention coming from experts in transitional economies such as Anders Aslund, who argues in favor of Western intervention in Ukraine (Anders 2015). Based on the considerations in this paper, we suggest the role of the international community to be limited to: 1) facilitating the dissemination of free and unbiased information on market economies and strong democracies in the world to the isolated populations in conflict or conflict prone areas, and 2) assisting the governments of countries like Ukraine to implement structural political and economic reforms informed by the theory and practice of successful institutional transformations in history, as well as by the local practices and knowledge.